\documentclass[journal]{IEEEtran}

\ifCLASSINFOpdf
   \usepackage[pdftex]{graphicx}
\else
  \usepackage[dvips]{graphicx}
\fi
\begin{document}
\title{A Technical Framework for\\ Musical Biofeedback in Stroke Rehabilitation}

\author{Prithvi~Kantan,~\IEEEmembership{}
        Erika~G.~Spaich,~\IEEEmembership{}
        and~Sofia~Dahl~\IEEEmembership{}
\thanks{
This paper was submitted for review on 1 Dec 2020
The work is partially funded by NordForsk's Nordic
University Hub, Nordic Sound and Music Computing Network NordicSMC, project number 86892.

P. Kantan and S. Dahl are  with the Department
of Architecture, Design and Media Technology, Aalborg University, Copenhagen, Denmark. E-mail: {prka, sof}@create.aau.dk}
\thanks{E. Spaich is with the Department of Health Science and Technology, Aalborg University, Aalborg, Denmark. E-mail: espaich@hst.aau.dk}
}

\markboth{}
{Kantan \MakeLowercase{\textit{et al.}}: A Technical Framework for Musical Biofeedback in Stroke Rehabilitation}

\maketitle

\begin{abstract}
We here present work  a generalized low-level technical framework aimed to provide musical biofeedback in post-stroke balance and gait rehabilitation, built by an iterative user-centered process. The framework comprises  wireless wearable inertial sensors and a software interface developed using inexpensive and open-source tools. The interface enables layered and adjustable music synthesis, real-time control over biofeedback parameters in several training modes, and extensive supplementary functionality. We evaluated the system in terms of technical performance, finding that the system has sufficiently low loop delay ($\sim$90 ms), good sensor range ($>$9 m) and low computational load even in its most demanding operation mode. In a series of expert interviews, selected training interactions using the system were deemed by clinicians to be meaningful and relevant to clinical protocols with comprehensible feedback (albeit sometimes unpleasant or disturbing) for a wide patient demographic. Future studies will focus on using this framework with real patients to both develop the interactions further and measure their effects during therapy.
\end{abstract}


\section{Background}\label{sec:introduction}

Stroke survivors commonly suffer physical deficits that manifest as disturbances to balance and gait \cite{strkDef2013}. 
Advances in affordable computer power and portable motion-sensing technology \cite{bmbf_Bal3} have led to an increasing role of technology in rehabilitation \cite{roleSensors}, for instance with biofeedback, where physiological or biomechanical information is made available to conscious experience to allow for greater self-awareness of these states, and modification where necessary \cite{bmbf_Book}.
\emph{Biomechanical biofeedback (BMBF)} \cite{bmbf_Giggins} based on bodily kinematics or kinetics is the type of biofeedback most directly applicable to neurorehabilitation, specifically balance/mobility as well as lower limb activities and gait\cite{bmbf_Bal1, bmbf_Gait2}. Results based on studies with both healthy and impaired populations indicate the superiority of biofeedback in training compared to regular therapy protocols in improving postural sway \cite{costantini,bmbf_Bal5}, weight shifting and reaction time \cite{bmbf_Bal1}, as well as sit-stand transfers \cite{bmbf_Bal6} and gait kinematics \cite{bmbf_Bal5,bmbf_Gait1}.

\emph{Auditory biofeedback (ABF)} involves the real-time conversion of measured bodily information into a sonic representation. By definition, it can thus be seen as a specific case of interactive sonification \cite{soniHandbook}, where data relations are rapidly converted into auditory relations \cite{abf_Bal2,costantini,abf_Gait6}. The supplied auditory information on movement execution serves as continuous or discrete guidance which assists in movement error correction \cite{strkRehab_G14,parseihianDyn,matsubara}.




\subsection{Auditory Biofeedback in Movement Training}

ABF has been applied to train postural control, with positive results \cite{abf_Bal2, costantini, strkBal_N3, abf_Bal5}.
In a series of studies, Dozza and colleagues explored the use of multidimensional ABF using a system that sonified trunk accelerations/sway velocities \emph{continuously} through frequency, level and spatial balance of a stereo sound using nonlinear mappings.
The biofeedback information provided was similar to that provided by the vestibular system \cite{dozza2011}, and the biofeedback improved balance overall, more so when other key sensory cues were unreliable or absent \cite{abf_Bal2}. 
Direction specificity of audio biofeedback reduced postural sway and increased the frequency of postural corrections in the direction of the biofeedback \cite{strkBal_N3,dozza2011}. Furthermore, the optimal mapping function for trunk sway to ABF was found to be sigmoid-shaped \cite{abf_Bal5}. 

Unlike the above continuous mapping paradigm, Costantini et al. \cite{costantini} successfully tested a biofeedback system that projected trunk inclination onto discrete 2D zones in the horizontal plane. Postural deviations triggered auditory warnings of proportional intensity using simple filtered and modulated noise.
Though the authors only performed short-term evaluations with unimpaired subjects, they found significant reductions in postural sway in several conditions \cite{costantini}.

Engardt et al. \cite{bmbf_Bal4} assessed the effects of ABF during training of \emph{sit-to-stand (STS)} transfers for hemiparetic stroke patients, finding short term improvements in body weight distribution between the paretic and non-paretic limb. Nicolai et al. \cite{bmbf_Bal6} found significant and sustained improvements in posture and balance post-intervention on patients with progressive supranuclear palsy. Patients received an auditory cue to stand up when an individually calibrated trunk flexion angle threshold was crossed \cite{bmbf_Bal6}.

ABF has also shown positive effects in gait training \cite{bmbf_Bal3,bmbf_Gait2,bmbf_Gait1}. 
For instance, sonifying ankle rollover patterns as a series of data-driven synthesizers was found to bring about significant differences in cadence and walking velocity among participants \cite{abf_Gait6}. Torres et al. \cite{abf_Gait4} introduced an \emph{inertial measurement unit (IMU)}-based prototype and prescribed a number of movement-sound couplings, such as fixed movement thresholds to trigger discrete auditory feedback or modulate continuous auditory feedback \cite{abf_Gait4}.

\subsection{Dynamic Trajectory Tracking}

The above systems essentially provide \emph{error-based} feedback \cite{strkRehab_G14}, where the difference between a quantity and a \emph{constant} ``target" value is sonified over time. 
However, in the context of a variable target (as in dynamic movement training), there is evidence that error feedback may not be most ideal in terms of performance outcomes \cite{parseihianDyn,matsubara}, although available research is inconclusive. 
Rosati et al. \cite{rosati} showed that error feedback did not improve performance w.r.t. visual feedback alone, while an auditory representation of the visualized target motion (\emph{task}-related feedback) was more valuable.  
However, Boyer et al. \cite{boyer} found that both these feedback types could reduce tracking error and increase movement energy in visuo-manual tracking. 
Parseihian et al. \cite{parseihianDyn} conducted an audio-guided 2D dynamic trajectory-tracking experiment based on the above research, and found that information related to what the user ``needs to do" resulted in superior tracking performance compared to error feedback. Due to the dynamic nature of the task, they found that pitch and other auditory dimensions that allow rapid comprehension and adjustments on the part of the user were most optimal \cite{parseihianDyn}.

The timing of task-related feedback presentation may also be critical, specifically whether task information is provided simultaneously with user feedback or slightly in advance (allowing for user anticipation).
An example of the former \cite{matsubara} concurrently presented two auditory streams corresponding to the task (reference) and user's own performance, panned to opposite stereo locations, with the user's goal being to make them sound identical. 
While the interaction was generally feasible and comprehensible, position- and timing-based user performance errors (relative to target) were found to be significantly worse than with visual feedback.
On the other hand, Parseihian et al. \cite{parseihianDyn} found that feedback based on \emph{anticipated distance error} afforded far superior performance to merely instantaneous distance error. 

Despite promise, ABF has failed to attain widespread practical adoption \cite{soni_Guide8,soni_Guide7}, partly due to a lack of focus on aesthetics and naturalness in sonic interaction design \cite{soni_Guide8} leading to poor user experience \cite{soni_Guide6}. Most ABF systems reviewed here, provide feedback through simple audio manipulations (e.g. pitch, loudness, brightness, spatialization), which generate aesthetically simple feedback signals. These are known to cause auditory fatigue, annoyance and dissatisfaction, making them less likely to be accepted by users \cite{soni_Guide8,vickers2016soniMusic,soni_Guide6,soni_Guide7,vogt2009physiosonic}.
Naturalness and clear causality in the iconic gesture-sound mapping of auditory displays have been found to contribute to their perceived usability \cite{soni_Guide11}, in line with the general aesthetics-usability correlation seen in human-computer interaction literature \cite{HCI_Aesth2}.


\subsection{Musical Biofeedback}
The recent exploration of \emph{musical biofeedback (MBF)}  \cite{music_Bf1,music_Bf12,soni_Guide13} has attempted to address the aesthetics issues of ABF and leverage the universal emotional and sociocultural appeal of music. MBF is a relatively complex signal due to its organization in time and frequency, possibly containing several coordinated instrumental and vocal elements that can provide variety to the feedback signal \cite{music_Bf12}. A general criticism levelled against such aesthetic approaches to sonification is that the interpretation of the underlying data is more difficult \cite{soni_Guide7,soni_Guide6}, and in the case of music entails the learning of a new `sonic grammar' \cite{musiSoniVickers}. It has, however, been argued that there is a cultural or aesthetic baseline in popular music systems, which is accessible to untrained listeners and allows them to appreciate music with minimal cognitive overhead in the absence of formal training \cite{musiSoniVickers}. For instance, listeners are able to recognize music genres within a fraction of a second \cite{mace_genre_2012}. The psychological and therapeutic benefits of music are well known \cite{sihvonen_music-based_2017} and  decades of research in the discipline of neurologic music therapy have established the direct therapeutic benefits of music across multiple dimensions in physical rehabilitation \cite{nmt_1}. 

A prescribed and largely prevalent MBF approach is to sonify desired movement behaviors as pleasant auditory states and vice versa, often simultaneously using musical rhythm to temporally organize motor timing \cite{music_Bf1}. The design space for possible interactions is conceivably vast and as such, MBF systems to date have ranged widely in scope and complexity, manipulating either pre-recorded musical stimuli or real-time synthesized ones as follows:

\paragraph{Pre-recorded Music}
Some studies have performed simple manipulations of existing music waveforms, for example by adding noise \cite{music_Bf23}, filtering \cite{music_Bf22,vogt2009physiosonic} or adjusting audio quality \cite{music_Bf18,vogt2009physiosonic} to sonify physiological and biomechanical quantities. These interactions were found to be comprehensible by healthy and impaired populations, and capable of positively altering motor behavior while reducing perceived exertion \cite{music_Bf22}.
Others sonified motor timing through music timing, such as the D-Jogger \cite{music_Bf9}, which synchronizes pre-existing music to detected gait patterns by time stretching algorithms, thus providing a sense of rhythmic agency to the user \cite{music_Bf19}.

\paragraph{Synthesized Music}
Other studies employed real-time synthesis approaches, through which it is easier to exert finer control over sonic parameters of music and more easily craft intimate and engaging musical interactions \cite{fabiani_InteractiveMusicSystems}. 
Sonification parameters used in these designs include musical pitch \cite{abf_Gait2,music_Bf27,music_Bf29}, tempo \cite{music_Bf12}, brightness \cite{music_Bf27}, track volumes \cite{music_Bf27}, chord arpeggio characteristics \cite{music_Bf13}, musical layer richness \cite{music_Bf14}, synthetic tone additions \cite{music_Bf14} and percussive sample triggering \cite{music_Bf25}. In most cases, the systems only underwent preliminary evaluation such as brief usability tests with convenience-sampled participants. 
But at the very least, the results indicate that these MBF interactions are feasible, perceptible and comprehensible, as well as potentially pleasurable experiences.

\subsection{Appraisal of Earlier Work}
Combining music with the portability, versatility and movement  modification of ABF can enable powerful mediation of human behavior \cite{music_Bf19}, since music can motivate, monitor and modify human movement \cite{music_Bf1}, and is as effective as simple sine sonification while reducing auditory fatigue \cite{music_Bf12}.
Hovever, the musical stimuli generated are usually rigid and simplistic, either monophonic instruments \cite{abf_Gait2} or very basic ensembles \cite{music_Bf14,music_Bf28}. The synthesis of stimuli resembling professionally produced music is undoubtedly challenging, and bare-bones aesthetics \cite{music_Bf13,vickers2016soniMusic} lacking the consideration of user preferences \cite{soni_Guide8,music_Fam} can hamper user experience. Systems also tend to be designed for specific applications, making their designs hard to generalize to other scenarios. While some works mention details of individual feedback tailoring to patients \cite{vogt2009physiosonic,abf_Gait4,linnhoffGait}, MBF literature typically does not provide detailed system design specifications, and data mapping configurations seem to be empirically designed and rigid, difficult to alter in real-time or retroactively tune \cite{soniHandbook} as part of user-centered approaches \cite{music_Bf19}. 

The use of expensive proprietary/custom-built hardware and software \cite{abf_Bal5,strkBal_N3,costantini,abf_Bal8,bmbf_Bal6,vogt2009physiosonic} makes these works difficult for other researchers to replicate and upgrade. Moreover, the use of visual programming environments in many studies \cite{abf_Gait3,music_Bf25,music_Bf9,music_Bf28,music_Bf23}, while excellent for preliminary testing, is computationally less efficient \cite{soni_Guide13} and arguably harder to scale in complexity than low-level programming languages, although technical performance details are seldom reported in research. E.g. most gait ABF systems claim to be `real-time', while few report feedback loop delay values \cite{linnhoffGait}. Lastly, it is unclear whether most systems possess the recommended BMBF supplementary functions \cite{bmbf_Book}, such as visualization, standby modes and time-series logging. 

The aim of the present work is to address several of the shortcomings in MBF system and build a clinically applicable MBF framework for balance and gait training using exclusively open-source development tools and an user-centred methodology \cite{higginbottom_participatory_2015}. As identified from the literature review, the system should fulfil a number of requirements such as 1) be computationally efficient, with adequate temporal performance and wireless sensing range; 2) have an architecture versatile enough to provide multiple types of user-adjustable training interactions within a single interface; and 3) these must be conceptually and practically applicable as a supplement to conventional stroke rehabilitation protocols.

We developed a musical biofeedback system in collaboration with patients and clinical stakeholders. The main contribution lies in the flexibility of the biofeedback mapping architecture, layered, adjustable and efficient music synthesis as well as its ease of replication due to open-source development tools.





\section{Design and Implementation} \label{sec:design}





\begin{figure} [thb]
    \centering
    \includegraphics[width = 1\linewidth]{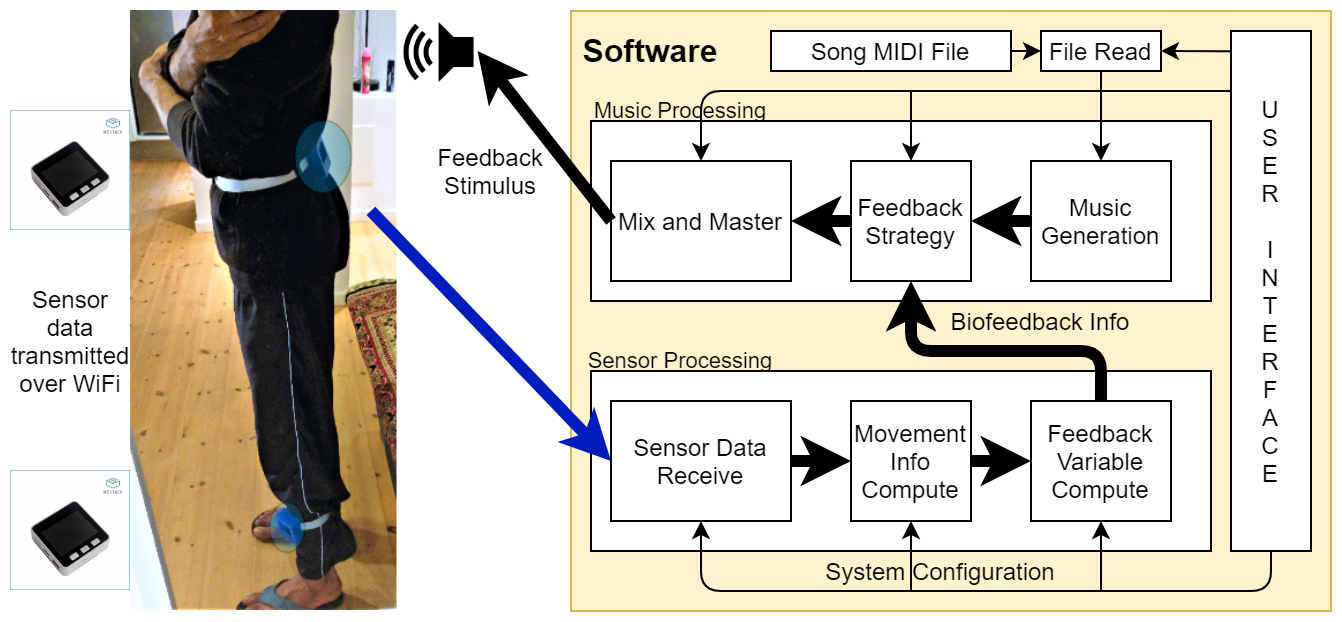}
    \caption{High-level system schematic showing the organization of the hardware and software components of the framework, and where the user is situated among them.}
    \label{fig:contextDiag}
\end{figure}


Architecturally, we opted for a \emph{distributed} structure \cite{bmbf_Book} with wearable wireless IMU's and remote processing on a laptop. Sensor interfacing, music generation and biofeedback configuration are controlled by a Windows application that produces a stereo audio signal, which is fed to the patient via headphones or loudspeakers as shown in Fig. \ref{fig:contextDiag}. The source code of the system is licenced under GNU GPL 3.0\footnote{https://github.com/prithviKantanAAU/mbfFramework\textunderscore v4}.

The hardware sensing component consists of M5Stack Grey microcontrollers that are programmed in the Arduino IDE. IMU data is transmitted to the software application as OSC \emph{(Open Sound Control)} messages over WiFi-UDP. We wrote the software component of the system in C++ using the JUCE environment\footnote{JUCE Framework - https://juce.com/}, which we chose for its extensive and efficient set of libraries for timer callbacks, OSC, MIDI, graphical elements and file operations. For music synthesis and sonification, we implemented a FAUST \footnote{FAUST Programming Language - https://faust.grame.fr/} script, which was compiled to generate a JUCE-compatible DSP object in C++ and thus directly leverage the vast audio DSP functionality of FAUST. The interface layout is organized into three tabs for sensor interfacing, music control and biofeedback control (Supplementary Material 1).

\paragraph{Sensing and Wireless Communication} Up to three M5Stack devices, securely mounted to the patient's lower back or ankles using a silicone housing and elastic straps, connect to a secure WiFi network hosted by the laptop via a bidirectional IP verification process. The sensors each transmit IMU data and battery status at 125 Hz to a predefined remote UDP port. The software constantly checks for new OSC messages received at each port to infer whether that sensor is online. The sensor tab in the software handles sensor assignment to body parts (trunk or either leg) along with bias calibration.

\begin{figure} [thb]
    \centering
    \includegraphics[width = 0.65\linewidth]{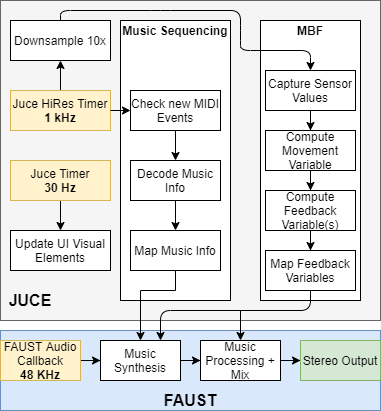}
    \caption{High-level software schematic showing how the different functions (UI update, music sequencing, biofeedback computation and music generation) are organized to run in timed callbacks at different frequencies.}
    \label{fig:callbackStruct}
\end{figure}

\paragraph{Software Application Topology}

The software can be seen as a multi-instrument music synthesizer (FAUST DSP object) that is `played' by the JUCE component, and this `performance' comprises \emph{sequenced music} from song files and \emph{MBF} based on the sensors, which manipulates the sequenced music. The FAUST object has virtual controls to trigger the instruments, modify their properties, configure music mixing processors and apply MBF strategies. The JUCE component handles all this in appropriately timed callback functions as shown in Fig. \ref{fig:callbackStruct}. A single high resolution timer orchestrates the music sequencing callback (at 1 kHz) and the MBF callback (at 100 Hz). The sensor transmission rate (125 Hz) exceeds the MBF callback frequency to compensate for UDP packet drops. User interface element update occurs independently at 30 Hz, while the real-time audio callback itself is handled by the FAUST object at a sampling rate of 48 kHz with a software buffer of 480 samples (adding 10 ms of software latency). The UI controls modify the states of sequencing and MBF variables using listener callbacks for thread safety.





\subsection{Music Generation}



The system generates an eight-track stereo instrumental ensemble containing both melodic and percussive elements in a 4/4 time signature. These fulfill musical `roles' corresponding to percussion, melody and harmony in a simplified pop music style. A music piece can be reproduced in various rhythmic grooves and instrument textures to cater to varied tastes, switchable in real-time in the interface. Other aspects of the music can also be varied on the fly, such as tempo, number of instruments, track balance and mix processing parameters. The generation is MIDI-based with the option to load external song files, although inbuilt music is also provided for specific training settings. The MIDI files are encoded in a custom Type-1 schema for efficiency, and generation process involves two separate processes - sequencing and synthesis.

\paragraph{Sequencing} Here, MIDI messages for all tracks are decoded into pitch and velocity information to map to the appropriate FAUST synthesizer controls. Song-specific information related to melody, bass and chord pitch is stored in loadable MIDI \emph{song files}. Instrument choices, rhythmic information and articulation for all tracks is encoded in four-bar \emph{style files} that are dynamically pre-populated at startup, making it easy to add new rhythms and styles to the software. 

MIDI information is stored as note matrices in program memory. During playback, the sequencing callback at 1 kHz (see middle branch of Fig. \ref{fig:callbackStruct}) increments the sequencer's elapsed MIDI ticks as per the configured tempo, checks MIDI timestamps in the note matrices for new events to be handled and counts them. The event types are identified (note on/off) and the event details (pitch/velocity) are preprocessed and mapped to the respective FAUST controls. The tempo slider controls playback rate by changing the tick increment per callback interval. Polyphonic tracks may have up to four voices (chords), and note frequencies are constrained to specific registers for pitched tracks to reduce sonic disparities among songs in different musical keys. Playback proceeds and the rhythmic pattern loops until all song events have been handled.

\paragraph{Synthesis} Audio generation uses a hybrid approach where percussion instruments are sample-based, while melody instruments are synthesized using custom-built algorithms and/or FAUST library functions where available. Each instrument role can be reproduced in up to three sonic \emph{variants}, and appropriate combinations of variants are used for each musical style preset. For example, the `hi-hat' percussive role can be played using a regular hi-hat, ride cymbal or marimba sample. Synthesis is mainly custom FM or subtractive \cite{thesis}, with basic physical models such as Karplus-Strong and formant-filtered vocal simulations. Pitch parameters influence note frequencies of pitched tracks, while velocity influences level and note articulation properties. Tracks have their own channel compressors and 4-band fully parametric equalizers with pre-defined but modifiable settings for each variant. Temporal parameters like envelope time-constants, reverberation and echo time automatically adapt to the tempo. The tracks are summed and undergo master processing (see Fig. \ref{fig:callbackStruct}) comprising a master equalizer and limiter, akin to standard mixing workflows.

\subsection{Movement-Music Interactions}

    \label{fig:zoneVisualizer}

Various MBF interactions are possible for gait and balance training, each employing suitable feedback strategies to convey performance information. The strategies were designed and tested for perceptual salience, with most of them internally employing divergent (one-many) mappings between data and audio parameters for perceptual magnification \cite{soniHandbook}. The feedback is 1D or 2D and usually error-based \cite{strkRehab_G14}, where a measured movement parameter is compared to a \emph{target value range}, and feedback is provided based on patient compliance (see Supplementary Material 5 for video links). 

\paragraph{Static Balance}
The principle is to reward the maintenance of a target trunk orientation (upright or otherwise) with pleasant music, and provide negative MBF proportional to deviations from this target. Angular trunk tilt from the vertical is projected onto the plane formed by the (\textit{mediolateral (ML) and anteroposterior (AP)}) axes, and allocated to one of six discrete feedback zones such that MBF intensity increases with distance from the configured target trunk orientation. The zones are concentric circular or elliptical ring shapes around the target as described in \cite{costantini}, with two rectangular zones to the extreme left and right. We made it possible to offset the target to a non-upright position and change its size to accommodate differently able patients. Strategies such as music dissonance, disturbance tones and ambulance siren \cite{thesis} may be used as negative feedback, the last of which provides helpful L-R directional cues \cite{strkBal_N3} through stereo panning.
\paragraph{Dynamic Balance}
Trunk training is common to stroke rehabilitation due to its strong contribution to balance and mobility recovery, and typically involves core stability exercises based on trunk flexion, extension, weight shift and reaching movements \cite{trunkTraining}. We augmented these exercises with the use of MBF as follows. For \textbf{Reaching}, 1D ML or AP trunk angle is sonified as melodic scale degree, allowing the patient to `play' melodies on an otherwise monotonous inbuilt music piece by tilting their trunk forward or sideways as part of a reaching motion. 
Melodic scale and key can be changed in real-time to mitigate interaction monotony.

For \textbf{Trunk Control} (flexion and extension), we extended the static balance interaction by modulating the 2D target to follow a horizontal plane trajectory that the patient must match, in a configurable linear, diagonal, circular, square or rhombic shape. Frequency and phase of the trajectory progress are music-synced so that rhythmic cues thereof can assist movement planning, and the frequency can be adjusted to a sub-multiple of the music tempo to suit the patient. Feedback can be discrete error-based (zones as in static balance), 2D task-based (ML and AP target position sonified as separate MBF strategies independent of the patient) or continuous anticipated distance error-based as in \cite{parseihianDyn}. In the last, a sum of sigmoid functions \cite{abf_Bal5} is used to map directional feedback \cite{strkBal_N3}. For rapid comprehension \cite{parseihianDyn}, an exemplar MBF configuration uses L-R spatialization for ML axis feedback and and melody pitch skewing for AP. 

\paragraph{STS}
As reviewed in \cite{balasubramanian2015analysis}, movements which happen in a continual fashion without any interruptions characterize most well-trained and healthy motor behavior, and serve as a marker of post-stroke motor recovery. This interaction rewards smooth sit-stand transitions with pleasant music, providing negative MBF to detected movement intermittencies by introducing salient disturbances in the music related to \textbf{jerkiness}. Mean-squared jerk captures instantaneous intermittencies with sufficient speed and sensitivity for meaningful smoothness-based feedback as no segmentation or windowing is required. 
\textbf{Trunk Flexion-based Cues} based on \cite{bmbf_Bal6} form the second interaction, where a sit/stand action cue is provided based on \textit{trunk flexion angle}. We augmented the principle  musically here, as a neutral sonic artifact (e.g. bell or wah wah effect \cite{thesis}) is triggered within the music when separately configurable sit/stand angle thresholds are exceeded.

\paragraph{Gait} The purpose of these interactions is to augment rhythmic auditory stimulation training \cite{nmt_1,sihvonen_music-based_2017} with synchronization-related feedback. The two interactions focus on step duration and phase respectively \cite{thesis}. In the former, the system detects every step taken by the patient and compares its duration to the musical beat interval, providing immediate negative feedback proportional to how much longer or shorter the step was than the beat interval (with a configurable dead-zone). The phase-based interaction is inspired by \cite{music_Bf25,music_Bf28}, where percussive musical events are triggered by footfalls, and the goal is to synchronize these events with the remaining ensemble, to encourage movement-music synchronization. Here, the bass drum and snare drum tracks are muted from the sequenced music and respectively triggered by the left and right foot. By walking in time, the patient is thus rewarded with a well-synchronized ensemble.

\subsection{Feedback Calculation and Mapping}

\begin{figure} [thb]
    \centering
    \includegraphics[width = 0.8\linewidth]{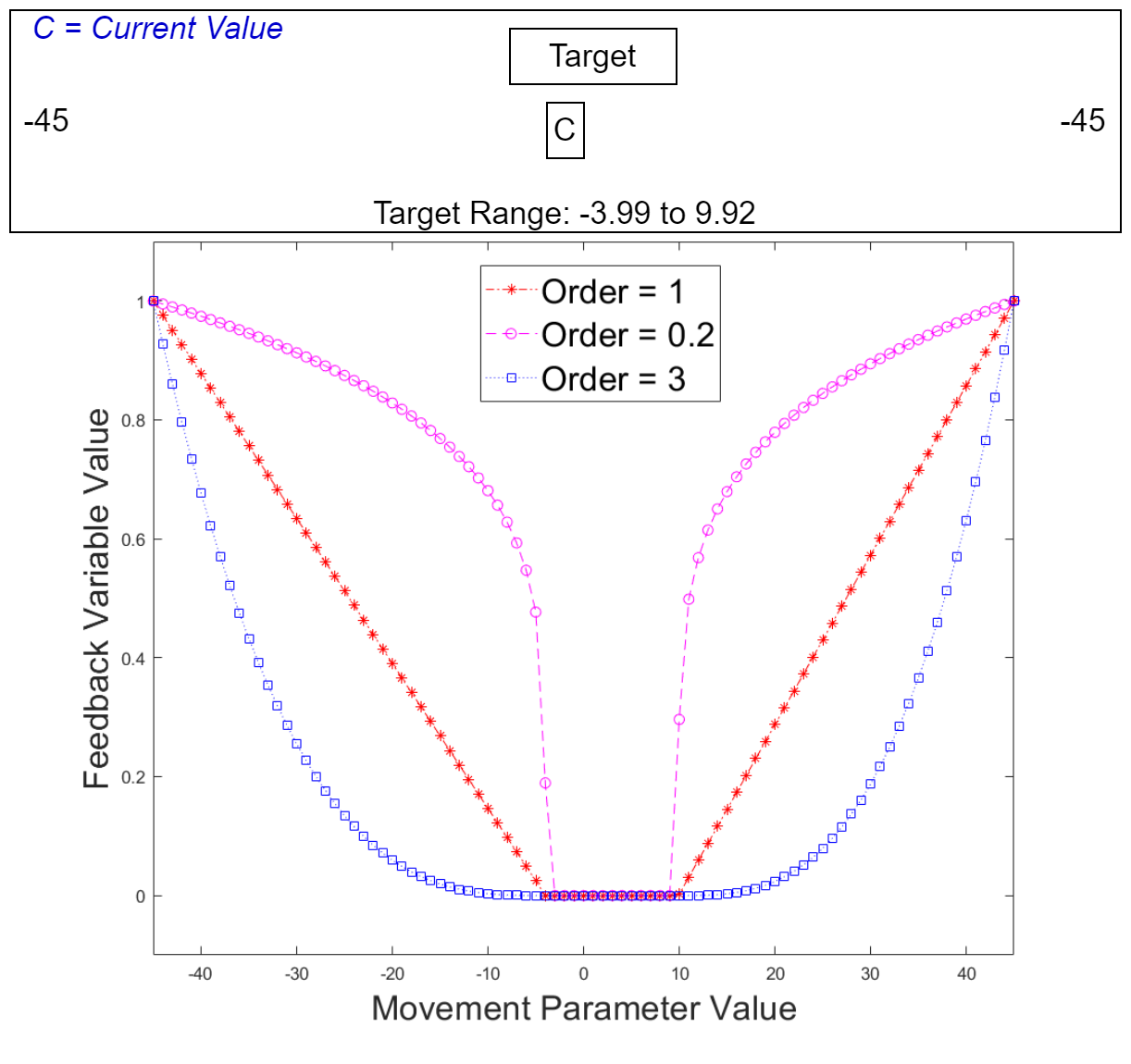}
    \caption{The 1-D MBF mapping function for a specific movement parameter target range to generate a feedback variable between 0 and 1. MBF intensity increases differently outside this range depending on the order of the gamma function.}
    \label{fig:1DBMBF}
\end{figure}

The above interactions require a dedicated functional block to transform IMU signals to meaningful feedback signals (rightmost branch of Fig. \ref{fig:callbackStruct}). The transformation must be flexible enough to be tailored to various patient types. Our framework allows any 1D mapping combination between computable movement parameters \cite{thesis} and the available MBF strategies, with extensive control of mapping parameters (interactive data selection and mapping \cite{soniHandbook}). The 100~Hz mapping frequency yields a perceptually smooth result.

The received raw IMU data first undergoes signal conditioning \cite{soniHandbook} - median filtering and smoothing using 6th order Butterworth lowpass filters. Filter parameters (median filter length in samples and cutoff frequency in Hz) are user-adjustable with suitable inbuilt defaults for each measured parameter. One of several exercise modes may be chosen (e.g. static balance, gait, etc.), each of which has its own set of movement parameters, MBF strategies and specific user controls. The movement parameter once computed is then transformed to a \emph{feedback variable}, which involves comparing the movement parameter value to the target range and normalizing compliance error between 0-1 \cite{soni_Guide8} followed by a gamma (exponential) function, which modifies the mapping function shape \cite{soni_Guide13} (see Fig. \ref{fig:1DBMBF}).

As shown in Figure~\ref{fig:1DBMBF}, a feedback variable value of 1 indicates maximum MBF intensity and vice versa. The interface allows real-time mapping function control \cite{soni_Guide13}, feedback variable quantization and polarity inversion \cite{soniHandbook}. The result is mapped to the chosen MBF control of the FAUST object, which accordingly manipulates the music output to provide biofeedback. For directional MBF strategies, a feedback variable value of 0.5 corresponds to no feedback, while 0 and 1 correspond to the directional extremes of feedback intensity.



\subsection{Supplementary Functionality}

The system allows real-time data visualization and session logging.
We implemented generalized graphical visualizers to monitor measured movement parameters. 
A progress bar shows song completion. Movement repetition data is also captured for dynamic reaching, STS and gait interactions. A time-series logging functionality is provided to capture training session progress in granular detail (100~Hz resolution) for further analysis. The system configuration state for a session can also be saved for future recall. Lastly, a standby mode option is provided to toggle biofeedback on and off to facilitate effect comparisons in future studies.

\section{Evaluation} \label{sec:evaluation}
The system and interactions were evaluated through technical tests (Supplementary Material 2) and expert interviews (see Supplementary Materials 3 and 4 for specifics).
\subsection{Technical Testing}
All tests were carried out on a Dell Inspiron 15 7000 Windows laptop with a 1.8 GHz i7 processor and 16 GB RAM.

\emph{Biofeedback loop delay} was measured by comparing onset timestamps between system input (movement) and output (sound) for trunk angle,  foot strike and jerk. 
Mean and standard deviations (in parenthesis) were
\textbf{90 (5) ms} for trunk angle and \textbf{93 (48) ms} for foot strike/jerk.

\emph{Computational performance} was measured as \% processor time, and logged using performance benchmarking software.
We found the computational worst case scenario to be the \textit{slow Rock style @ 60 BPM, dynamic balance interaction with session logging on}.
The measured \% Processor Time was \textbf{28.91 (4.09) (Peak: 40.80)}. Peak CPU usage (Windows Task Manager) was \textbf{11.1 \%} along with 157.0 MB of memory.

\emph{Effective indoor sensor range} was assessed in terms of the percentage of MBF callbacks that detected new OSC messages in a short time-frame under different conditions. Results showed MBF Callbacks with new OSC data received up to 96.35 \% for 3 m direct line of sight; 96.10 \% for 7 m direct line of sight; and 82.50\% for 9 m around wall corner.
 
\subsection{Expert Interviews }
In addition to involving patients and a lead physiotherapist in the design iterations, we assessed the interactions from the perspective of clinicians and patients via structured expert interviews conducted online with five physiotherapists and two music therapists after the third development cycle. Details of interview materials, questionnaires, data analysis and results are in Supplementary Materials 3 and 4, and a short summary is given here.

The participants stated that the most of the interactions were applicable to a range of acute and sub-acute patient types depending on the complexity of the training activity. They also expressed that the available software adjustments would be sufficient to tailor the training to different individuals. A recurring comment about the gait interactions was that they are more applicable to cerebellar or lower brainstem strokes than the more common cortical strokes. The participants estimated that the sensing mechanism would capture relevant movement features in all training activities, and that the auditory feedback in some cases would provide them with some extra information not as readily available visually (e.g. jerkiness, gait rhythm).

The experts also felt that the interactions could both be integrated into existing clinical protocols and used to enhance patient autonomy when not undergoing training. For example, several suggested that the static balance interactions could be used to provide continuous feedback while performing other tasks. Numerous suggestions to adapt the interactions for other goal-oriented training activities (e.g. upper limb) were also made. While the participants found the majority of musical feedback strategies easily perceptible and comprehensible, they commented on the unpleasant nature of some of them, particularly those involving salient synthetic disturbances to the music. In contrast, strategies that operated on the level of musical structure (eg. foot strike drum triggering, music stop outside target range, music pitch) were seen more positively. Several also felt that the music sounded quite computerized, although individual patient reactions would vary. In terms of practicality, most felt that the sensor setup was simple, although one stated that there could be hygiene issues with using velcro straps. Several brought up patient safety concerns during training, stressing that the therapist needed to be able to operate the system in a hands-free manner.

\section{Discussion and Conclusions} \label{sec:discussion} 

We presented an MBF framework for post-stroke movement rehabilitation co-developed with stakeholders and addressing several shortcomings of existing systems by integrating theories of BMBF system design \cite{bmbf_Book}, auditory guidance \cite{soni_Guide8}, music therapy \cite{nmt_1}, musical biofeedback \cite{music_Bf1} and interactive sonification \cite{soniHandbook,soni_Guide13}. Our framework enables multiple interactions catering to conventional protocols for balance and gait training within a single hardware-software architecture with granular real-time system control. Though tests of the final system with real patients are pending due to COVID-19 restrictions, our evaluation found the interactions to be technically sufficient, applicable to a wide demographic and relevant to therapy protocols.

The fact that our system was built using free open-source tools and easily available, cheap hardware, also makes it accessible to the research community for replication and improvement. The ESP32-OSC-JUCE-FAUST approach yielded a practically feasible prototype. We found the sensing hardware to have sufficient range, and the software performed very efficiently even in its computational worst-case scenario, with overall system loop delays well below the human auditory reaction time \cite{bmbf_Book,linnhoffGait}. This approach makes high-level FAUST functionality available in a low-level programming environment, combining the rapid prototyping advantage of the former with the robustness and efficiency of the latter. The available computational and memory headroom allows future versions to be adapted for mobile versions and scaled in complexity when real-life tests expose necessary upgrades.

All interactions and most MBF strategies were deemed by the participating experts to be meaningful and easily comprehensible for patients. This supports our argumentation for the universal comprehensibility of musical meaning without formal training \cite{musiSoniVickers}. The experts pointed out the excessive unpleasantness of some negative MBF strategies, although real-life tests with patients must further investigate this trade-off between feedback salience and pleasantness to inform an optimal MBF design philosophy. As reviewed in \cite{linnhoffGait}, positive feedback may generally be more conducive to long-term motor learning than negative feedback as it promotes motivation and invokes dopamine prediction error mechanisms as opposed to simply attentive processing of movement errors. Future studies will focus the use of our generic mapping framework on providing positive feedback. Overall, the technical framework also can be used with other patient groups than stroke by merely adding movement parameters and MBF strategies.

 The present functionality does make an effort to address the aesthetics problem \cite{soni_Guide8,soni_Guide7} and accommodate user preferences \cite{music_Fam}, but the music generation still has significant upgrade potential. From the comments made by the clinicians (not primarily music experts), there seem to be clear aesthetic limitations. The synthesis methods are relatively simple, and the sequencing process is deterministic and predictable with limited temporal variability. Even taking into account subjectivity in music preferences, the system generally produces a computerized-sounding output. Future versions of the system could integrate computational rules \cite{fabiani_InteractiveMusicSystems} for expressive music performance to yield a more vibrant stimulus.

Although the expert interviews did provide insight, they cannot replace real usability tests and clinical effect studies to gauge therapeutic benefits from interacting with the system, which is the next step. Future work could also include comparisons of our system with other systems built using other tools and architectures. Nevertheless, the framework we have proposed and built has the potential to facilitate the future evaluation of various MBF paradigms in stroke rehabilitation.

\section{Acknowledgments} We thank Helle Rovsing Møller Jørgensen and the participating patients and therapists. Author PK was main responsible for system development and manuscript writing. Authors SD and EGS supervised the MSc project \cite{thesis}, and assisted in writing. All authors approved of the final manuscript.

\appendices

\IEEEpeerreviewmaketitle

\ifCLASSOPTIONcaptionsoff
  \newpage
\fi

\bibliography{bibtex/bib/IEEEabrv.bib,bibtex/bib/mybib.bib}{}
\bibliographystyle{IEEEtran}

\begin{IEEEbiography}
[{\includegraphics[width=1in,height=1.25in,clip,keepaspectratio] {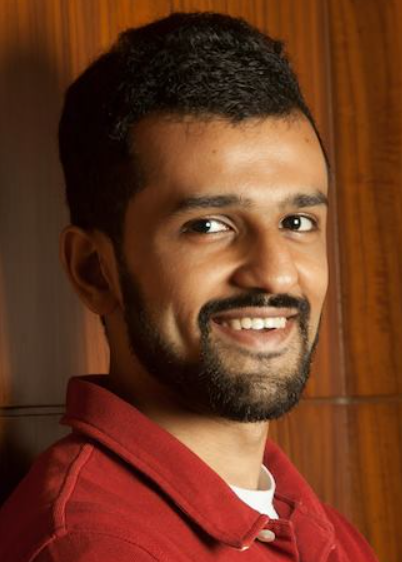}}]
{Prithvi Kantan}
holds a bachelor degree in Electronics and Telecommunications Engineering from Mumbai University and an MSc. in Sound and Music Computing from Aalborg University, Copenhagen, where he presently works as a research assistant. His primary interest lies in the research and development of music technology for the healthcare domain, while he has also worked with rhythm perception and interactive musical sonification.
\end{IEEEbiography}

\begin{IEEEbiography}
[{\includegraphics[width=1in,height=1.25in,clip,keepaspectratio] {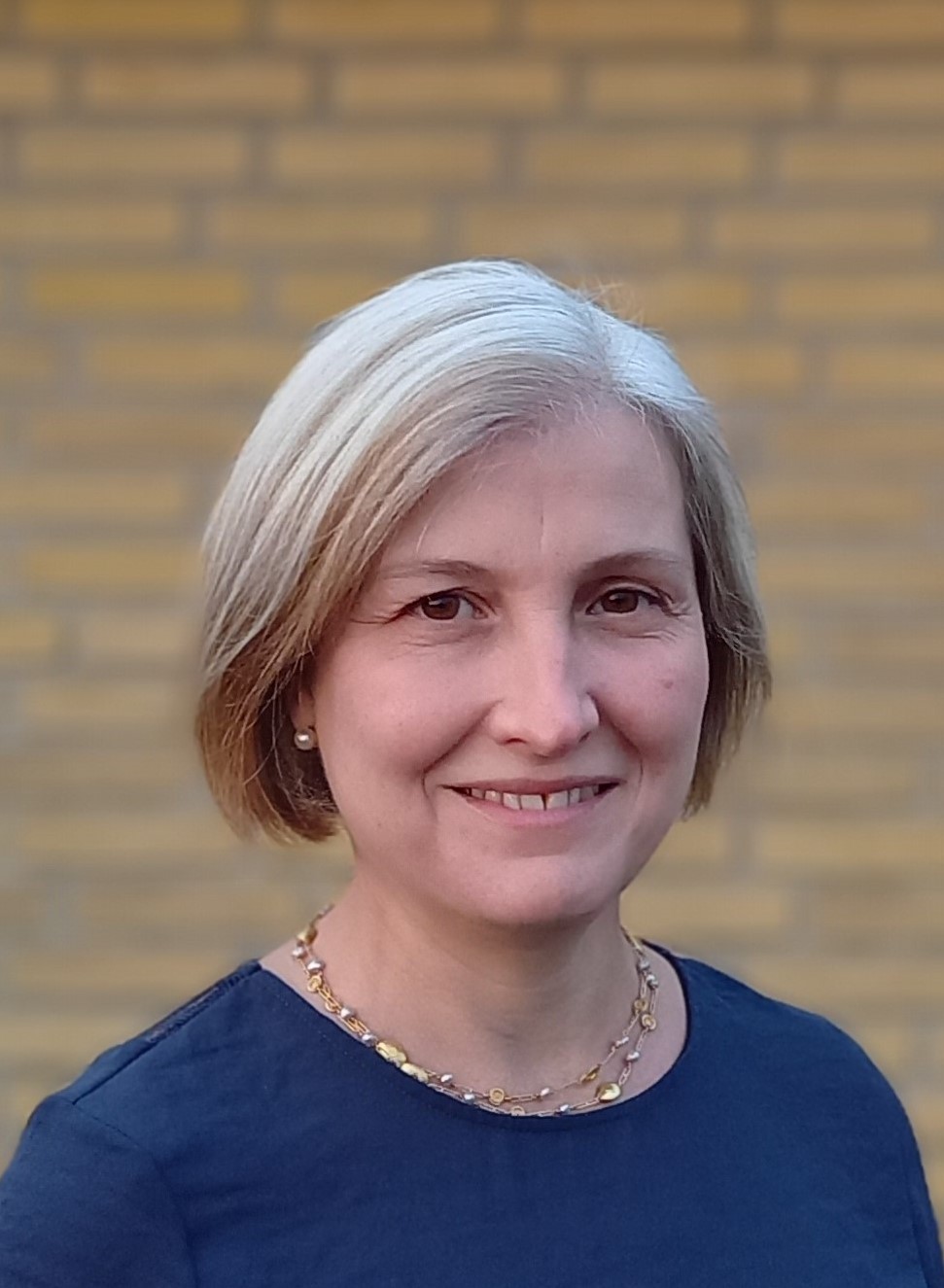}}]
{Erika G. Spaich}
received the Bioengineering degree from the National University of Entre Rios (UNER), Argentina, in 1998, and the Ph.d. degree in Biomedical Science and Engineering from Aalborg University, Denmark, in 2004. She is currently an Associate Professor at the Department of Health  Science and Technology, Aalborg University, Denmark. Her research interests include neurorehabilitation, functional electrical stimulation and therapy, characterization and use of the nociceptive withdrawal reflex in the rehabilitation of the hemiparetic gait, rehabilitation robotics, technologies for assessment of rehabilitation programs, and sensory-motor physiology. Dr. Spaich is the Vice-president of the International Functional Electrical Stimulation Society (IFESS).
\end{IEEEbiography}

\begin{IEEEbiography}
[{\includegraphics[width=1in,height=1.25in,clip,keepaspectratio] {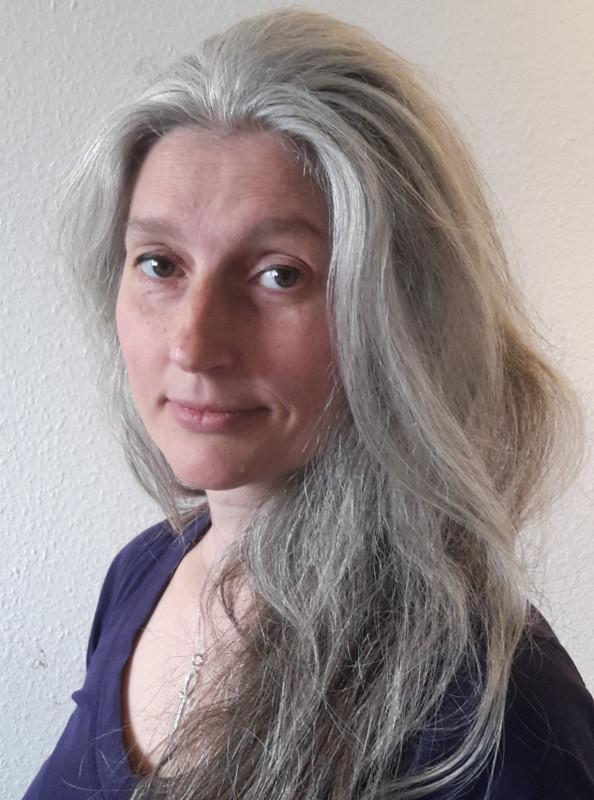}}]{Sofia Dahl}
holds a PhD in Speech and Music communication from KTH, Royal Institute of Technology, Sweden. As associate professor at Aalborg University, her primary research field is within embodied music cognition, but her work spans disciplines such as music technology and acoustics, psychology, neuroscience, and music performance. 
Dr. Dahl is in the steering committee for the Nordic Sound and Music Computing University Hub, funded by Nordforsk, and currently serving in the Executive Council of the European Society for the Cognitive Sciences of Music (ESCOM).
\end{IEEEbiography}

\end{document}